# *p*-process in SNIa

**Claudia Travaglio**
*INAF – Astronomical Observatory Turin*
*Strada Osservatorio 20, 10025 Pino Torinese, (Italy)*
*E-mail:* `travaglio@oato.inaf.it`

**Roberto Gallino[a], Wolfgang Hillebrandt[b], Friedrich Röpke[c,b]**
[a]*University of Turin, Physics Department, Via P. Giuria 1, 10125 Torino (Italy)*
[b]*Max-Planck- Institut für Astrophysik,Karl-Schwarzschild-Str. 1, D-85748 Garching (Germany)*
[c]*Universität Würzburg, Emil-Fischer-Str.31, D-97074 Würzburg, (Germany)*

We explore SNIa as *p*-process sources in the framework of two-dimensional SNIa models using enhanced *s*-seed distributions as directly obtained from a sequence of thermal pulse instabilities. The SNIa WD precursor is assumed to have reached the Chandrasekhar mass limit in a binary system by mass accretion from a giant/main sequence companion. We apply the tracer-particle method to reconstruct the nucleosynthesis from the thermal histories of Lagrangian particles, passively advected in the hydrodynamic calculations. For each particle we follow the explosive nucleosynthesis with a detailed nuclear reaction network. We select tracers within the typical temperature range for *p*-process production, $1.5\text{-}3.7\ 10^9$K, and analyse in detail their behaviour, exploring the influence of different *s*-process distributions on the *p*-process nucleosynthesis. We find that SNIa contribute to a large fraction of *p*-nuclei, both the light *p*-nuclei and the heavy-p nuclei at a quite flat average production factor. For the first time, the very abundant Ru and Mo *p*-isotopes are reproduced at the same level as the heavy *p*-nuclei. We investigate the metallicity effect on the *p*-process production. Starting with a range of *s*-seeds distributions obtained for different metallicities, running SNIa two-dimensional models and using a simple chemical evolution code, we give estimates of the SNIa contribution to the solar *p*-process composition. We find that SNIa contribute for at least 50% at the solar *p*-nuclei composition, in a primary way.

## 1. Introduction

The *p*-process was introduced by the pioneering works of Cameron (1957) and Burbidge et al. (1957), and was postulated to produce 35 stable isotopes between $^{74}$Se and $^{176}$Hg on the proton-rich side of the valley of stability. The largest fraction of *p*-isotopes is synthesized during γ-process by sequences of photodissociations and β-decays (see e.g., Woosley & Howard 1978; Rayet et al. 1990; Arnould & Goriely 2003). This has been suggested to occur in explosive O/Ne burning during Type II supernova (hereafter SNII) explosions, and can reproduce the solar abundances for heaviest *p*-isotopes within about a factor of 3 (see e.g., Rayet et al. 1995: Rauscher et al. 2002). More recently *p*-process nucleosynthesis calculations in SNeII have been performed by Hayakawa et al. (2006, 2008), Farouqui et al. (2009). A major failure of most of these models is that the solar system abundance pattern of light *p*-nuclei (in particular $^{78}$Kr, $^{92,94}$Mo, $^{96,98}$Ru) is not reproduced, and therefore it is not possible to reproduce the solar abundances of the *p*-isotopes by one single process. For these nuclei alternative processes and sites have been proposed by different authors, e.g., strong neutrino fluxes in the deepest layers of SNII ejecta (Frölich et al. 2006), ν*p*-process in neutrino driven winds of SNII (Woosley et al. 1990; Pruet et al. 2006; Roberts et al. 2010; Wanajo et al. 2006; Arcones & Janka 2011). Still, the yields of light *p*-nuclei normalized to the most abundant isotope from SNII, i.e. $^{16}$O, indicate the need of another important stellar source for them.

The *p*-process has also been suggested to occur in the outermost layers of Type Ia supernovae (hereafter SNIa), in the Chadrasekhar mass white dwarf delayed detonation model (Howard & Meyer 1993: Travaglio et al. 2011 hereafter TRV11), in the sub-Chadrasekhar mass white dwarf He-detonation model (Goriely et al. 2005; Arnould & Goriely 2006), and in the carbon deflagration model (Kusakabe et al. 2011). A fundamental role for the *p*-production is played by the *s*-process seed distribution assumed, and most of these authors discussed different hypotheses for the *s*-process seeds distribution. Again, as for SNII, for most of these works the solar system abundance pattern of light *p*-nuclei is not reproduced. Up to now, only TRV11 demonstrated that multidimensional models of delayed detonation SNIa, accurately following the outermost part of the star, can reproduce both light *p*-nuclei below *A*=120 and the heavy-p nuclei in comparable amounts with respect to $^{56}$Fe including $^{78}$Kr, $^{92,94}$Mo, and $^{96,98}$Ru. Note that 2/3 of the solar abundance of $^{56}$Fe is attributed to SNIa ejecta. In this work we use a standard delayed detonation model called DDT-a, and we investigate in detail different metallicity distributions for *s*-process seeds. Making use of a simple chemical evolution code (used by Travaglio and collaborators to study the role of AGB stars at different metallicities in chemical evolution of our Galaxy and their contribution to the Solar System composition, see Travaglio et al. 1999, 2001, 2004), we explore the role of *s*-process seeds at different metallicites and the contribution to the solar abundances of *p*-nuclei. Finally, we analyze the problem of the long-lived $^{92}$Nb and $^{146}$Sm. These two nuclides are shielded against the *s*-process as well as against the r-process. Measurements in meteoritic material firmly established the presence of these two isotopes at the formation of the solar system (see e.g., Harper & Jacobsen 1992; Yin et al. 2000; Dauphas et al. 2002.

## 2. SNIa hydrodynamic models

We follow the SNIa explosion model described in TRV11. The scenario is based on the explosion triggered once the carbon+oxygen white dwarf (hereafter CO-WD) approaches the Chandrasekhar mass. Thermonuclear burning starts out as a subsonic deflagration and later turns into a supersonic detonation. We refer to the single-degenerate scenario in which the white dwarf accretes material from a main-sequence or evolved companion star. The explosion is modelled using a two-dimensional hydrodynamic simulations following the delayed detonation model DDT-a, shown in Figure 1.

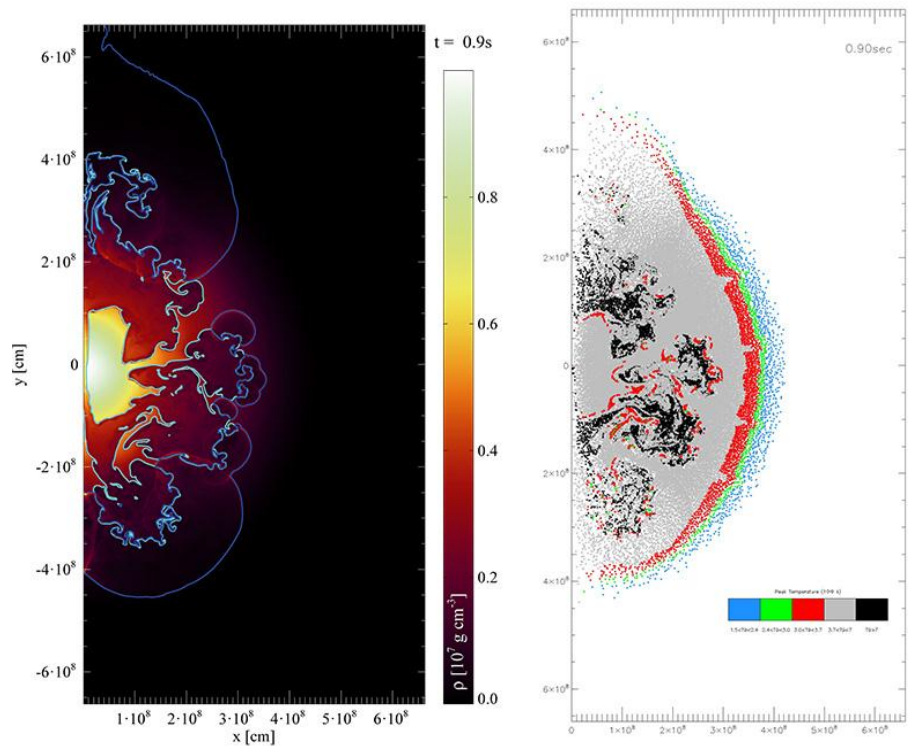

**Figure 1.** Snapshot from model DDT-a at 0.95 sec after ignition. On the left, the hydrodynamic evolution is illustrated by color-coded density and the locations of the deflagration flame (cyan contour) and the detonation front (blue contour). On the right-hand side the tracer distribution is shown. Wherease the locations correspond to the current time, the color coding is according to the maximum temperature reached during the explosion: black tracers peak with $T_9^{peak} > 7.0$; gray tracers with $3.7 < T_9^{peak} < 7.0$; tracers marked in blue ($1.5 < T_9^{peak} < 2.4$), green ($2.4 < T_9^{peak} < 3.0$), and red ($3.0 < T_9^{peak} < 3.7$) are peak temperatures reached in ranges where the *p*-process nucleosynthesis is possible.

## 3. Nucleosynthesis with tracer particles and *s*-process seeds

Since only a crude description of the thermonuclear burning is applied in the hydrodynamic explosion simulation, details on the nucleosynthesis are recovered in a postprocessing step. A Lagrangian component in the form of tracer particles is introduced over the Eulerian grid in order to follow and store the temperature and density evolution of the fluid.
We consider the tracers with a very small mass (Travaglio et al. 2004 and Seitenzahl et al. 2010 tested that the mass used for those particles give a reasonably good resolution). Nevertheless it is impossible (due to limited computational resources) to consider mixing processes between tracers, in particular for the slower tracers in the outermost part of the star (responsible for the p-process nucleosynthesis) this is certainly a very good approximation.
For the DDT-a model we used 51,200 tracer particles, uniformly distributed over the star, each one with a mass of about $3.0 \times 10^{-5}$ $M_\odot$. During the hydrodynamic simulation for each tracer particle $T$ and $\rho$ histories are recorded along their paths. The nuclear post-processing calculations are then performed separately for each particle.

The *p*-process nucleosynthesis is calculated using a nuclear network with 1024 species from neutron and proton up to $^{209}$Bi combined with neutron, proton, and $\alpha$-induced reactions and their inverse. The code used for this work was originally developed and presented by Thielemann et al. (1996). We use the nuclear reaction rates based on the experimental values and the Hauser–Feshbach statistical model NON-SMOKER (Rauscher & Thielemann 2000). Theoretical and experimental electron capture and $\beta$-decay rates are from Langanke & Martínez-Pinedo (2000).

The *p*-process nucleosynthesis occurs in SNe of Type Ia starting on a pre-explosive *s*-process enrichment; therefore, it is essential to determine the *s*-process enrichment in the exploding WD. We assume recurrent He-flashes occurring in the He-shell during the accretion phase. The matter accumulated onto the CO-WD therefore becomes enriched of *s*-nuclei, where neutrons are mainly released by the $^{13}$C$(\alpha,n)^{16}$O (Iben 1981; TRV11). This applies under the assumption that a small amount of protons are ingested in the top layers of the He intershell. Protons are captured by the abundant $^{12}$C and converted into $^{13}$C via $^{12}$C$(p,\gamma)^{13}$N$(\beta^+\nu)^{13}$C at $T \sim 1\times 10^8$ K.

We recall that the mass involved and the profile of the $^{13}$C-pocket still has to be treated as a free parameter. Note that in our hyphotesis the *s*-process enrichment occurs during the accretion process (therefore in a completely different environment than $^{13}$C-pocket formation during AGB phase). Since no real model are available at the moment for the production of *s*-seeds in accretion phase, we explore different *s*-seed distribution in order to better understand the dependence of our results on these initial seeds.

This process enables the synthesis of *p*-nuclei during the explosive phase. In TRV11 are analyzed and discussed different *s*-seed distributions, with consequences for the synthesis of *p*-nuclei. We fixed the $^{13}$C-pocket strength (Gallino et al. 1998) and we vary the metallicity (we explore $Z = Z_\odot$, 0.01, 0.006, 0.003). The *s*-process seeds production factors, normalized to solar $X_i/X_{i\odot}$, in the accreted matter are shown in Figure 2. A flat distribution for the *s*-only isotopes (black dots) is obtained at $Z = 0.01$, like in the Arlandini et al. "stellar model" (1999), although with mass fractions $X_i$ typically a factor of 2 higher. Decreasing the metallicity, the *s*-process distribution gets progressively distorted from a flat distribution in correspondence of neutron magic numbers N=50 (around A=90), at N=82 (around A=140) and eventually at N=126, at the termination of the *s*-process path (around A=208).

Note that all neutron magic nuclei are not classified as *s*-only, although they are predominantly of *s*-process origin.

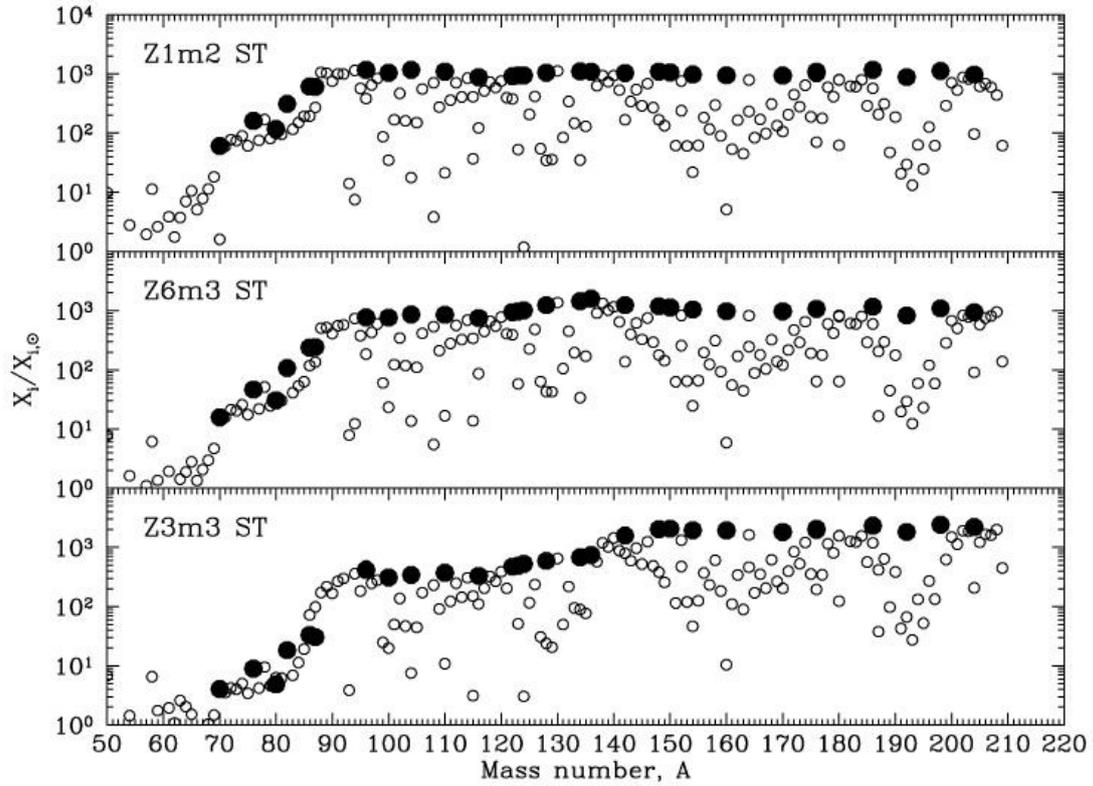

**Figure 2.** Distribution of initial seed abundances relative to solar for $Z = 0.01$ (upper panel), 0.006 middle panel), 0.003 (lower panel), ST case (see text).

## 4. Results

In Figure 3, we plot the production factor of each isotope with respect to solar normalized to the ratio of $^{56}$Fe obtained by the model for different metallicities (Z=0.01, 0.006, 0.003). We note that for many of the *p*-isotopes the overproduction is at the level of $^{56}$Fe.

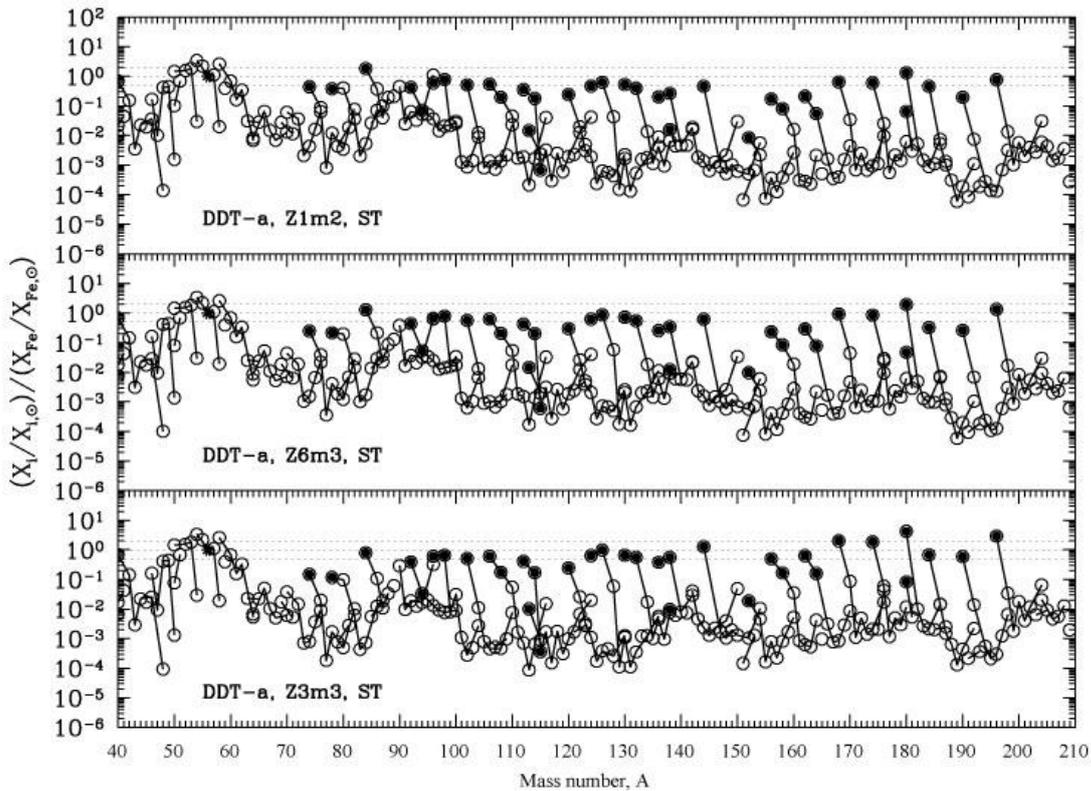

**Figura 3.** Nucleosynthesis yields (production factors normalized to Fe) obtained using 51.200 tracer particles in the two-dimensional DDT-a model (as described in the text). The *s*-process enrichment has been considered in the accreted mass with Z=0.01 (upper panel), 0.006 (middle panel) and 0.003 (lower panel).

Starting from the lightest *p*-isotopes, $^{74}$Se, $^{78}$Kr, $^{92}$Mo, $^{96,98}$Ru, $^{102}$Pd, and $^{106,108}$Cd are produced at the level of $^{56}$Fe (within a factor of two). In the case of $^{84}$Sr and $^{94}$Mo we are aware of large uncertainties in the theoretical predictions of the MACS at $kT = 100$ KeV (the typical temperature for explosive conditions), following Rauscher & Thielemann (2000). We verified that small changes in the rate of the main chain producing these isotopes change their abundances by a large factor. These nuclear uncertainties have to be carefully taken into account in *p*-process nucleosynthesis calculations. The results are clearly progenitor dependent. We did the calculations for Chandrasekhar-mass DDT models investigating different strength of detonation, as well as for pure deflagration models. This is explained in detail by TRV11. But one should also look at alternative scenarios, such as WD-WD mergers, where we expect a marginal *p*-process nucleosynthesis.

Note that, decreasing metallicity, the *s*-only seed isotopes (as shown in Figure 2) do not maintain a Solar System distribution. The resulting *p*-distribution remains consistently flat.
A more detailed work is in preparation, where we will analyze different $^{13}$C-pocket and *s*-process seed distributions and consequences for the *p*-nuclei produced in SNIa.

We conclude that taking a fixed choice of the $^{13}$C-pocket at different Z, we may infer that the $(p/^{56}Fe)/(p/^{56}Fe)_\odot$ ratio is not so much dependent on metallicity. This suggests a primary nature of *p*-process.

## 5. Chemical evolution

As far as the Galactic evolution of *p*-nuclei is concerned, we used the Galactic chemical evolution code used for different works presented by Travaglio and collaborators (Travaglio et al. 1999, 2001, 2004). For this work we extended the matrix of the isotopes, including the *p*-nuclei and we follow their evolution over time/metallicity up to solar metallicity. We include in the code the *p*-nuclei abundances at various metallicities discussed in the previous Section, making use of a fine interpolation among them. Taking into account the role of SNIa alone, we obtained the result shown in Figure 4.

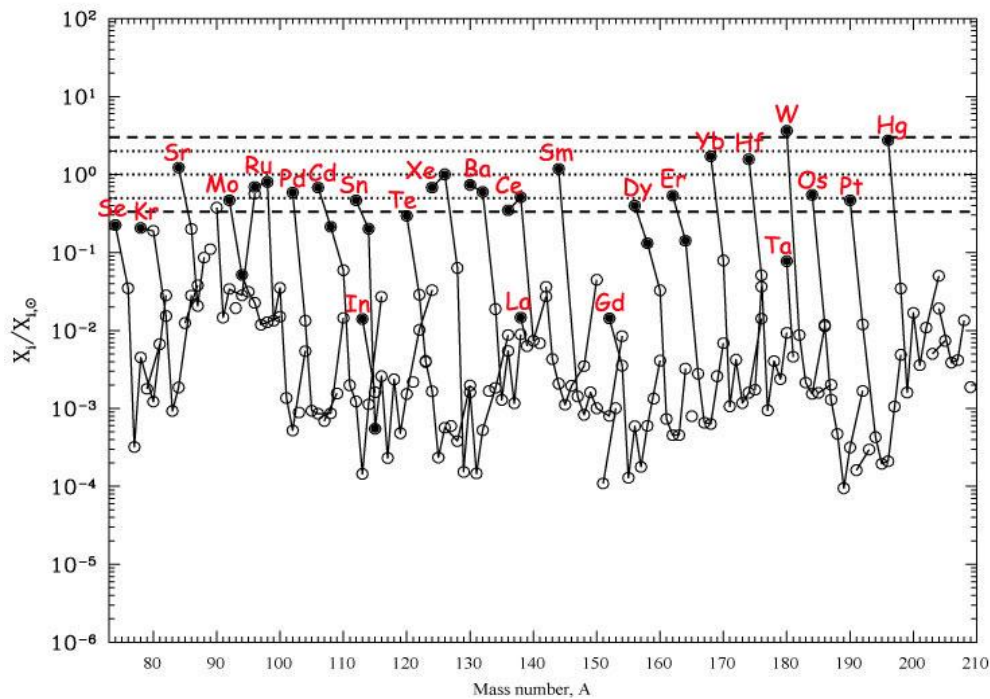

**Figura 4.** Chemical evolution results at solar composition (normalized to solar)

As one can see in Figure 4, few nuclei originally ascribed to the *p*-only group ($^{113}$In, $^{115}$Sn, $^{138}$La, $^{152}$Gd, and $^{180}$Ta) are far below the average of the other *p*-nuclei production. This is an indication for a different nucleosynthetic production for them. Concerning $^{113}$In and $^{115}$Sn, as discussed by Dillmann et al. (2008) and Nemeth et al. (1994), they can get important contributions from *β*-delayed r-process decay chains.

As for $^{138}$La, Woosley et al. (1990) demonstrated that the ν-nucleus interaction can contribute appreciably to the synthesis of $^{138}$La in the neon shell of SNII. $^{152}$Gd instead is predominantly of *s*-process origin, as it was demonstrated by Arlandini et al. (1999) and Käppeler et al. (2011). Like $^{152}$Gd, also $^{164}$Er is of predominant *s*-process origin, driven by the β-decay channel at $^{163}$Dy, which becomes unstable at stellar temperature (Takahashi and Yokoi 1987). As to the *p*-only nuclide $^{158}$Dy, its relative low abundance should be analyzed in the framework of present nuclear uncertainties of the γ-proceses. Finally, $^{180}$Ta, as discussed by TRV11 and Mohr et al. (2007), receives an important contribution from the *s*-process due to the branching at $^{179}$Hf, a stable isotope that becomes unstable at stellar temperatures. Also the ν-nucleus interaction can substantially feed $^{180}$Ta (Mohr et al. 2007).

Other two important isotopes need to be discussed here. The two lightest Mo isotopes ($^{92}$Mo and $^{94}$Mo) are still open issues. By TRV11 as well as in this work, $^{92}$Mo is well reproduced together with all the other *p*-nuclei. Instead $^{94}$Mo is still puzzling. As demonstrated by Howard et al. 1991, most of the $^{94}$Mo is produced from $^{98}$Mo through (γ,n) chain reactions.

Under the hypothesis that SNIa are responsible for 2/3 of the solar $^{56}$Fe, and assuming that our DDT-a model represents the typical SNIa with a frequency of 70% (Li et al. 2011), we conclude that they can be responsible for at least 50% of the all *p*-nuclei. SNII are expected to give a further and important contribution to the solar abundances of *p*-nuclei.

## 6. *p*-process chronometers

Whereas $^{93}$Nb is 85% *s*-process and 15% r-process (Arlandini et al. 1999), $^{92}$Nb is a shielded *p*-only nuclide; $^{92}$Mo is also a *p*-only nuclide, and can be choosen as a stable reference isotope for further discussion. The value of the ratio $^{92}$Nb/$^{92}$Mo calculated in the interstellar medium at the time of the Solar System formation is $2.01 \times 10^{-3}$ (see e.g., Yin et al. 2000; Dauphas et al. 2002). Various calculations are available in literature of *p*-process yields of $^{92}$Nb and $^{92}$Mo, and thus the relative ratio, for both SNIa (Howard, Meyer & Woosley 1991; Howard & Meyer 1993) and SNII (Woosley & Howard 1978, 1990; Rayet et al. 1995; Hoffman et al. 1996) supernova sources. Most of these ratios are too low and would not explain the meteoritic values. Only the ratio of 0.65 from Hoffman et al. (1996) for SNII sources with neutrino-driven winds can explain the high $^{92}$Nb/$^{92}$Mo value observed in the Early Solar System. With our calculation we obtain a value of $1.3 \times 10^{-3}$ for the ratio $^{92}$Nb/$^{92}$Mo.

Also the longlived $^{146}$Sm was found to be present in the Early Solar System, with an estimated ratio $^{146}$Sm /$^{144}$Sm = 0.01 (Jacobsen & Wasserburg 1984). In our calculations we find 0.32 value for that ratio.

## 7. Conclusions

We have presented results of detailed *p*-process nucleosynthesis calculations for two-dimensional delayed detonation SNIa models. We used initial *s*-seeds created during the mass accretion phase (as described by TRV11), under the assumption that a small amount of protons are ingested at the top layers of the He intershell.
We calculated *p*-process nucleosynthesis for different metallicities and with the tool of a simple chemical evolution code, we gave estimates of the role of SNIa in the contribution of p-nuclei abundances at the Solar System composition. We concluded that p-nuclei seems to be of primary origin, and that SNIa can contribute for at least 50% to the solar abundance of p-nuclei.
A much more detailed study of these aspects will be examined in a forthcoming paper.

**Achnowledgments**
This work has been supported by B2FH Association. The numerical calculations have also been supported by Regione Lombardia and CILEA Consortium through a LISA Initiative (Laboratory for Interdisciplinary Advanced Simulation) 2010 grant.
The work of FKR is supported by the Emmy Noether Program of the Deutsche Forschungsgemeinschaft (RO 3676/1-1), by the ARCHES prize of the German Ministry of Education and Research, and by the Group of Eight/Deutscher Akademischer Austauschdienst (Go8/DAAD) German-Australian exchange program.